# Second-order temporal coherence of polariton lasers based on an atomically thin crystal in a microcavity

Hangyong Shan[1,†], Jens-Christian Drawer[1,†], Meng Sun[2,†], Carlos Anton-Solanas[5], Martin Esmann[1], Kentaro Yumigeta[6], Kenji Watanabe[7], Takashi Taniguchi[8], Sefaattin Tongay[6], Sven Höfling[4], Ivan Savenko[3,9,10,*] and Christian Schneider[1,*]

[1]Institute of Physics, Carl von Ossietzky University, 26129 Oldenburg, Germany.
[2]Faculty of Science, Beijing University of Technology, 100124 Beijing, China
[3]Guangdong Technion Israel Institute of Technology (GTIIT), 241 Daxue Road, Shantou, Guangdong Province 515603, P.R. China
[4]Julius-Maximilians-Universität Würzburg, Physikalisches Institut and Würzburg-Dresden Cluster of Excellence ct.qmat, Lehrstuhl für Technische Physik, Am Hubland, 97074 Würzburg, Deutschland.
[5]Depto. de Física de Materiales, Instituto Nicolás Cabrera, Instituto de Física de la Materia Condensada, Universidad Autónoma de Madrid, 28049 Madrid, Spain.
[6]School for Engineering of Matter, Transport, and Energy, Arizona State University, Tempe, Arizona 85287, USA
[7]Research Center for Electronic and Optical Materials, National Institute for Materials Science, 1-1 Namiki, Tsukuba 305-0044, Japan
[8]Research Center for Materials Nanoarchitectonics, National Institute for Materials Science, 1-1 Namiki, Tsukuba 305-0044, Japan
[9]Technion -- Israel Institute of Technology, 32000 Haifa, Israel
[10]Guangdong Provincial Key Laboratory of Materials and Technologies for Energy Conversion, Guangdong Technion—Israel Institute of Technology, Guangdong 515063, China

*Corresponding author. Email: christian.schneider@uni-oldenburg.de, ivan.g.savenko@gmail.com
†These authors contributed equally: Hangyong Shan, Jens-Christian Drawer, Meng Sun

## ABSTRACT

Bosonic condensation and lasing of exciton-polaritons in microcavities is a fascinating solid-state phenomenon. It provides a versatile platform to study out-of-equilibrium many-body physics and has recently appeared at the forefront of quantum technologies. Here, we study the photon statistics via the second-order temporal correlation function of polariton lasing emerging from an optical microcavity integrated with an atomically thin $MoSe_2$ crystal. Furthermore, we investigate the macroscopic polariton phase transition for varying excitation powers and temperatures. The lower-polariton exhibits photon bunching below the threshold, implying a dominant thermal distribution of the emission, while above the threshold, the second-order correlation transits towards unity, which evidences the formation of a coherent state. Our findings are in agreement with a microscopic numerical model, which explicitly includes scattering with phonons on the quantum level.

## INTRODUCTION

Bose–Einstein condensation (BEC) is a macroscopic, quantum-degenerate state in thermal equilibrium. It exhibits long-range spatial coherence and long-term temporal coherence. Below the critical temperature, particles undergo a phase transition and spontaneously transit into a collective, in-phase state. In 1995, it was first experimentally observed in ultra-cold atom systems, requiring ultra-low temperature close to absolute zero ($10^{-6}$ K) [1,2]. As an alternative to these technically demanding BEC platforms, solid-state systems offer several advantages in the study of BEC phenomenology such as higher working temperatures, simpler setups, and unique driven-dissipative characteristics.

Exciton-polaritons in microcavities, one kind of bosonic quasi-particles arising from the strong coupling between cavity photons and excitons, are an ideal solid-state platform to study out-of-equilibrium bosonic condensation and lasing [3-5]. Owing to the photonic component, the effective mass of polaritons is remarkably light, being as low as $10^{-4}$ of the electron mass. This character is an intrinsic, distinct advantage to facilitate condensates at higher temperatures, even at room-temperature [6-10], a *sine qua non* factor in wide-spread potential technology applications. Polariton condensation exhibits most of BEC features including the occurrence of spatiotemporal coherence [5], despite the fact that it is generally in dynamic quasi-equilibrium since in most cases, the lifetime of polaritons is shorter than their relaxation time [3].

Recently, the long-range spatial coherence of monolayer-based exciton polaritons has been reported, with the integration of excitons in atomically-thin transition metal dichalcogenide (TMDCs) crystals such as $MoSe_2$ [11], $WSe_2$ [12] and $WS_2$ [13,14] monolayers. Those monolayer semiconductors are emerging hosts of two-dimensional excitons, which present extraordinary properties. For instance, they have large exciton binding energies, on the order of hundreds of meVs [15], a prerequisite for polaritonics at room temperature.

Since a polariton condensate/lasing is a collective quantum matter wave, its interferometric wave property can be manifested by the first-order spatial correlation [4,16]. The second-order coherence, on the other hand, characterizes intensity fluctuations of the lasing and discloses statistics of photonic emission in the time domain. It is used to identify the quantum nature of a state [17] and to evaluate the feasibility of polariton lasing as a coherent light state with a shot-noise-limited intensity stability [18].

In this Letter, we study the second-order coherence of polariton lasing created from an atomically thin $MoSe_2$ crystal strongly coupled to a monolithic microcavity. We observe a characteristic intensity threshold in the nonlinear input-output curve of the polariton emission, as well as a linewidth narrowing, implying the occurrence of polariton lasing. Furthermore, we perform Hanbury Brown and Twiss (HBT) measurements, recording the second-order coherence as a function of pump power, to probe the polariton distribution along the input-output transition. Then, we identify the distinct behavior of the second-order coherence evolving with temperature, indicating the thermal depletion of the degenerate polariton state. Finally, we present a theoretical model based on the Lindblad master equation [19] to describe the dissipative dynamics of the system. Using the Monte Carlo wavefunction technique [20], we model the behavior of second-order coherence in the experiments.

## RESULTS

### Sample and exciton-polaritons lasing

Figure 1(a) shows a schematic diagram of the sample structure. It is composed of an AlAs /$Al_{0.2}Ga_{0.8}As$ bottom distributed Bragg reflector (DBR), a $MoSe_2$ monolayer (capped by a thin hexagonal boron nitride layer), a spin-coated polymethyl methacrylate (PMMA) layer acting as the microcavity spacer, and finally, a dielectric top DBR. The top mirror is fabricated by a micro-mechanical assembly technique, similar to the dry-transfer method of monolayers [21]. Precise parameters of the structure are introduced in the Refs. [11], [22]. The quality factor (Q-factor) of the resulting cavity is ~760 as assessed by white light reflection measurements (see the Supplemental Material (SM) [22] for more details).

Experiments are conducted at a cryogenic temperature of 3.2 K, the optical setup is introduced in the supplementary material [22]. In the r.h.s. panel of Fig. 1(b), we plot the angle-resolved white light reflectivity to reveal the dispersion relation. The spectrum consists of two separate resonance branches and the energy gap. It unambiguously verifies the formation of exciton-polaritons, originating from the strong coupling between excitons of the $MoSe_2$ monolayer and the cavity photons.

In Fig. 1(b), a classical two-coupled oscillator model is applied to fit the polariton dispersion relation. Herein, the eigenstates of upper (UP) and lower polaritons (LP) are expressed as

$$E_{UP,LP}(k_{\parallel}) = \frac{1}{2}\left(E_x + E_c - i(\gamma_x + \gamma_c) \pm \sqrt{4g^2 + (E_c - E_x - i(\gamma_x - \gamma_c))^2}\right),$$ where $g$ is the coupling strength,

$E_x$ and $\gamma_x$ are the exciton energy and dissipation rate, $E_c$ and $\gamma_c$ are the cavity photon energy and dissipation rate, $\gamma_x$ ($\gamma_c$) corresponds to the half widths at half-maximum of exciton (cavity photon) [23]. The energy splitting of UP and LP, obtained from the reflectivity data, is $11 \pm 1$ meV. In the fitting, $E_x$ (dotted line) and $E_c$ (dashed line) at $K_{\parallel} = 0$ μm$^{-1}$ are 1.6565 eV and 1.6465 eV, respectively. $2\gamma_c$ is 2.2 meV, extracted from the reflectivity of empty cavity in Fig. S2, and $2\gamma_x$ is homogeneous broadening of exciton spectral linewidth, assumed as 2 meV in accordance with four-wave mixing measurements in this reference [24]. See more details of the fit in the SM [22].

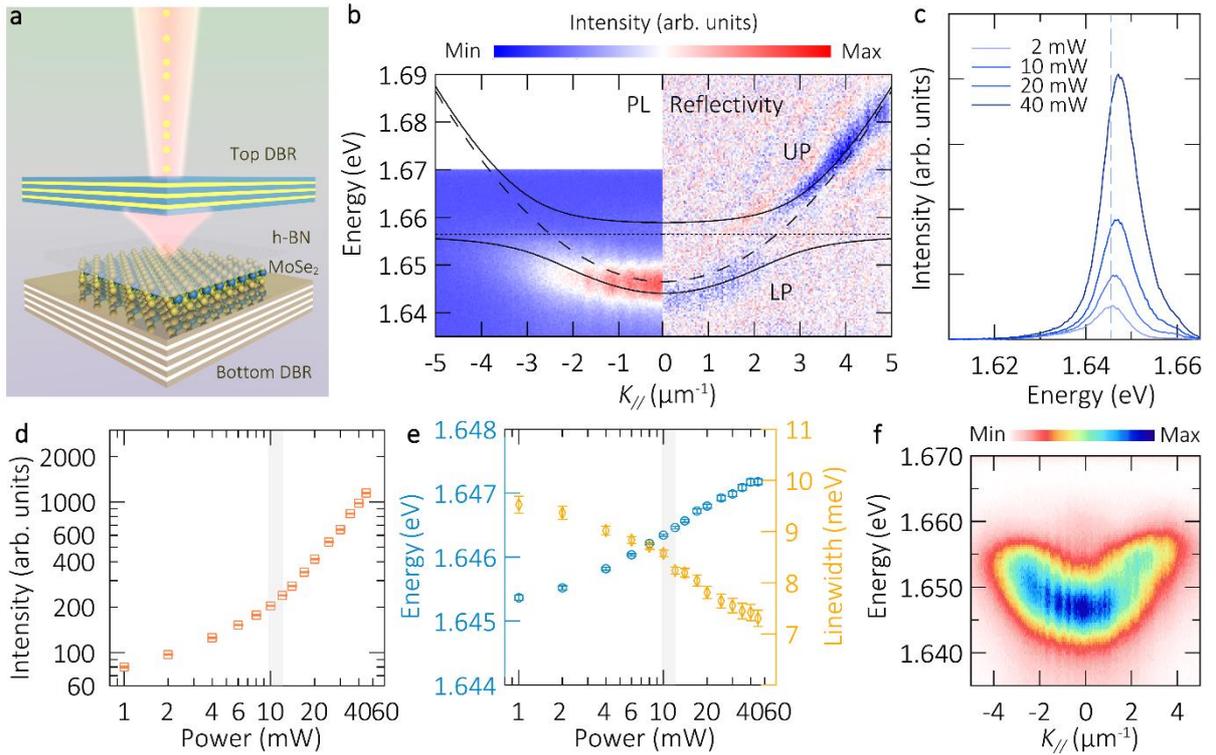

**Fig. 1 Sample structure and polariton lasing.** (a) Sketch of the sample structure. The optical cavity is composed of two DBRs. The MoSe$_2$ monolayer is transferred by dry-stamping technique and is capped by a thin layer of hBN. (b) Left: angle-resolved PL map. The energy of the pulsed excitation laser is set at 1.675 eV. Right: angle-resolved white light reflectivity. The fit depicts the dispersion relation of polaritons, extracted from the angle-resolved spectral data. The fitted dispersion relations of upper (UP) and lower polaritons (LP) are shown as solid lines. The dotted and dashed lines represent dispersions of bare excitons and microcavity photons, respectively. (c) Emission spectra of LP at around $K_{\parallel} = 0$ μm$^{-1}$. The pump power ranges from 2 to 40 mW. The dashed line denotes the spectral position of the LP peak under a 2 mW pump. (d) The integrated intensity of the LP emission versus pump power. The input-output results feature a kink at pump power of 10 mW, which is assigned to the threshold of lasing. The error bars are obtained from the standard deviation of the background noise. (e) Extracted LP linewidth and peak energy as a function of pump power. The peak energy (linewidth) is marked as blue (orange) dots. At 10 mW, a remarkable reduction in linewidth appears, and it is a fingerprint of the phase transition. The energy and linewidth error bars correspond to the 95% confidence interval of the peak fitting. (f) Angle-resolved PL map above the threshold at 20 mW.

Photons emitted from the radiative decay of polaritons allow us to directly probe polaritonic properties through optical measurements. We utilize angle-resolved photoluminescence (PL) to access the polariton occupancy at the ground state and to study polariton lasing. A linearly polarized picosecond pulsed laser is used as the excitation source, operating at 740 nm (1.675 eV) with a 2 ps pulse duration. It resonantly injects carriers into the upper polariton (UP) branch. The resulting angle-resolved PL map is displayed in the left panel of Fig. 1(b). It depicts a curved dispersive mode with an identical shape and energy as the reflectivity signal, confirming that this mode is exactly the LP state.

Polariton lasing occurs when the mean LP population exceeds unity, and the phase of stimulated polariton-polariton scattering is locked to the phase of the ground state. To assess this polariton lasing effect, we study the LP emission as a function of pump power. In Fig. 1(c), we plot the LP emission spectra at different pump powers. Those spectra are extracted from the angle-resolved PL map at $K_{//}{\sim}0$ µm$^{-1}$. We find the peak energy continuously blueshifts, and its linewidth is reduced as the laser power increases, which are two crucial features of polariton lasing.

Next, we fit the LP emission signals with a Lorentzian function to quantitatively evaluate the density-dependent behavior. Figure 1(d) shows the integrated LP emission intensity versus pump power, plotted in a double-logarithmic scale. In this input-output curve, the slope obviously increases and deviates from the linear region of emission intensity when the power exceeds 10 mW. Such a nonlinear emission feature can be assessed as a signature of the macroscopic accumulation of polaritons in the ground state [3,12].

In Fig. 1(e), we present the power-dependent evolution of the LP linewidth (orange) and peak energy (blue). Below the threshold, the linewidth slowly decreases with pump power, yet a notable drop is observed at 10 mW. It is indicative of an enhancement of phase coherence of the state, strongly supporting the occurrence of a polaritonic phase transition. Above the threshold, the linewidth continues to reduce, as the phase coherence of polaritons further improves at higher injection density [12]. Fig. 1(f) is the angle-resolved PL map above the threshold at 20 mW, the emission is dominated at around $K_{//}{=}0$ µm$^{-1}$. The slightly asymmetric dispersion is attributed to the geometry shape of the MoSe$_2$ flake, as well as its orientation relative to the slit of spectrometer.

The polariton-polariton repulsion, resulting from the exciton-exciton interaction and fermionic screening, gives rise to a dynamics blueshift throughout the entire polariton density range [25]. The blueshift magnitude (~2 meV) is much smaller than the mode-splitting energy, suggesting that the strong coupling regime is maintained over this pump power range.

Macroscopic spatial coherence is a crucial criterion for the claim of polariton phase transition. With this same device, and utilizing comparable pumping conditions, we have exhaustively investigated its first-order autocorrelation function and demonstrated spatial coherence extended up to 4 µm [11].

**Second-order temporal coherence of polariton lasing**

In this section, we focus on the second-order temporal coherence properties of polariton lasing, studied via the HBT experiment. The second-order correlation function $g^{(2)}(\tau)$ is expressed as

$$g^{(2)}(\tau) = \frac{<I(t)I(t+\tau)>}{<I(t)><I(t+\tau)>},\qquad(1)$$

where $I(t)$ and $I(t+\tau)$ are the emission intensities at times $t$ and $t+\tau$, respectively, and $\tau$ is the time delay. The symbol $<...>$ indicates the time average. The degree of second-order coherence $g^{(2)}(\tau)$ quantifies the intensity fluctuations of polariton emission, and the value at zero delay $g^{(2)}(0)$ is of most interest, since it distinguishes between classical $g^{(2)}(0){>}1$, coherent $g^{(2)}(0){=}1$, and sub-Poissonian $g^{(2)}(0){<}1$ emission statistics.

In our experiments, the excitation and detection conditions yield a time-averaging effect. In the following, we distinguish between the original, unintegrated second-order correlation function $g^{(2)}(\tau)$, and the measured, effectively time-averaged correlation function $\overline{g^{(2)}}(\tau)$.

Under pulsed excitation, the measured $\overline{g^{(2)}}(\tau)$ is an integration of the original $g^{(2)}(\tau)$ within the pulse duration $\Delta T$, i.e. 2 ps in our case. For a short integration time, $\Delta T \ll T_c$ ($T_c$ being the polariton coherence time); the measured $\overline{g^{(2)}}(0)$ does not substantially deviate from the original $g^{(2)}(0)$ value. However, in the opposite limit of a long integration time $\Delta T \gg T_c$, the magnitude of measured $\overline{g^{(2)}}(0)$ is greatly reduced, and it may even approach 1 due to the temporal averaging effect [26]. Regarding exciton-polaritons in MoSe$_2$ monolayers, their typical coherence time is expected to be on the order of hundreds of femtoseconds [11], which is one order of magnitude shorter than the laser pulse duration, i.e., in line with the latter situation.

First, we perform HBT measurements along the input-output curve. Figure 2(a) shows the second-order correlation function $\overline{g^{(2)}}(\tau)$ under pulsed excitation at 0.6P$_{th}$. We find $\overline{g^{(2)}}(0)$ = 1.020±0.005, similar to sub-threshold measurements of exciton-polaritons in other material systems [18,26]. Although $\overline{g^{(2)}}(0)$ is not much larger than unity, the correlation peak reaches above the noise level of side peaks, i.e. we observe photon bunching. This bunching demonstrates the thermal statistics of polariton emission below the threshold, as expected. For thermal sources, the expected $g^{(2)}(0)$ value is 2, while polaritons generally do not reach thermal equilibrium because of their short lifetimes, so the corresponding $g^{(2)}(0)$ ranges between 1 and 2 [27]. Furthermore, the measured $\overline{g^{(2)}}(0)$ is significantly affected by the excitation and detection conditions as a result of the aforementioned temporal averaging, thus the measured $\overline{g^{(2)}}(0)$ value does not exceed 1 by much.

When the power is increased to 3.5P$_{th}$, as shown in Fig. 2(b), $\overline{g^{(2)}}(0)$ is reduced to 1.004± 0.003, manifesting clear suppression of photon bunching. The peak value at zero delay is submerged into the noise of the side peaks. It implies that beyond the threshold, photon number fluctuations in the polariton system are reduced, proving the phase transition from a thermal to a coherent state.

The full dependence of $\overline{g^{(2)}}(0)$ on pump power is displayed in Fig. 2(c) (open circles). The value of $\overline{g^{(2)}}(0)$ is mostly above 1.01 below threshold power, and it experiences a notable decrease at the threshold, approaching unity with the power. We also plot the emission intensity of the uncorrelated peaks in HBT measurements versus pump power as squares in Fig. 2(c), which qualitatively mimics the results shown in Fig. 1(d) (regarding the nonlinear emission behavior with the threshold power of 10 mW).

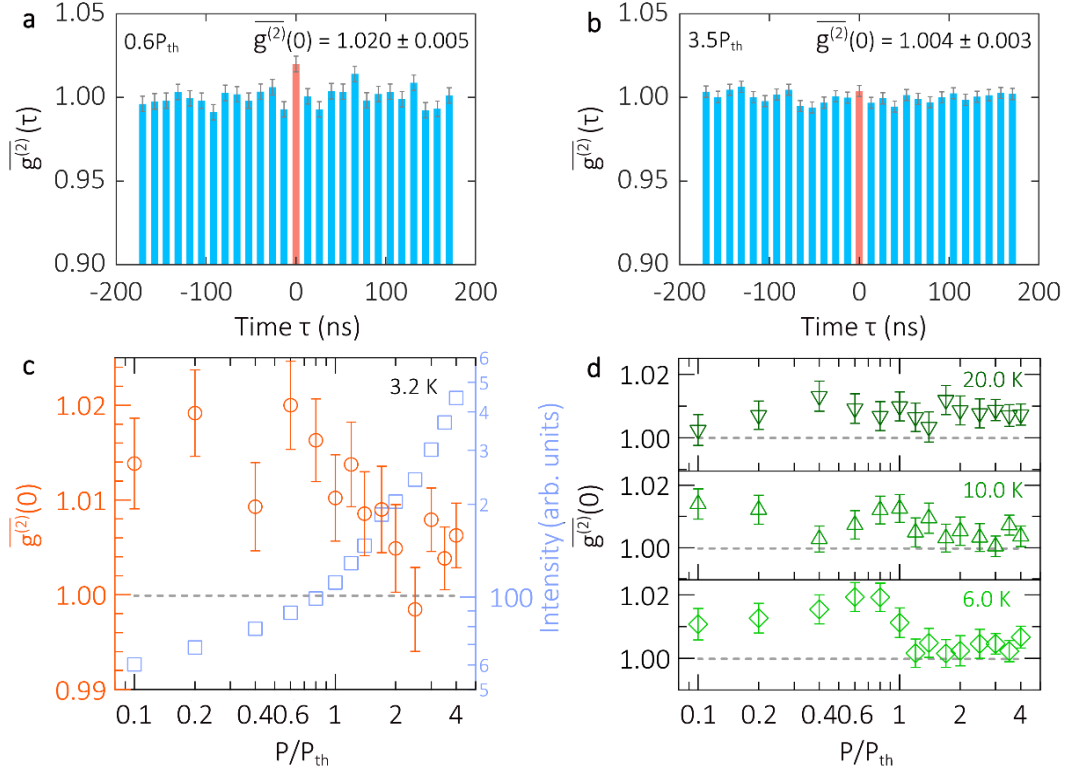

**Fig. 2 Second-order correlation function of polariton emission.** (a), (b) The second-order correlation function below [above] the polariton lasing threshold, 0.6P$_{th}$ [3.5P$_{th}$]. The value of $\overline{g^{(2)}(0)}$ is 1.020±0.005 [1.004±0.003], respectively. The temperature is 3.2 K. P$_{th}$ = 10 mW. (c) The second-order correlation function at zero delay $\overline{g^{(2)}(0)}$ (circles) and intensity of uncorrelated peaks (squares) as a function of pump power. The magnitude of $\overline{g^{(2)}(0)}$ experiences a reduction towards unity at threshold power. (d) $\overline{g^{(2)}(0)}$ as a function of pump power at temperatures of 6.0, 10.0 and 20.0 K (from bottom to top panels).

To theoretically model the behavior of the second-order coherence function, we apply the Lindblad master equation to describe the dissipative nature of the polariton system. Within the Born approximation, the total density matrix operator is $\rho_{tot} = \rho \otimes \rho_{pump} \otimes \rho_{ph}$. The terms on the r.h.s. of the equation are the density operator of polaritons, reservoir particles, and phonons, respectively. Partially tracing out the pumping and phonon degrees of freedom, we find that the evolution of the density matrix is given by

$$\frac{d\rho}{dt} = -\frac{i}{\hbar}[\hat{\mathcal{H}}, \rho] + \sum_s \left( \hat{J}_s \rho \hat{J}_s^\dagger - \frac{1}{2}\{\hat{J}_s^\dagger \hat{J}_s, \rho\} \right). \tag{2}$$

The first term describes the Hamiltonian dynamics of the system,

$$\hat{\mathcal{H}} = \sum_k E_k \hat{a}_k^\dagger \hat{a}_k + \sum_{k_1 k_2 p} U_{k_1 k_2 p} \hat{a}_{k_1}^\dagger \hat{a}_{k_2}^\dagger \hat{a}_{k_1+p} \hat{a}_{k_2-p}, \tag{3}$$

where $E_k$ is the lower polariton mode, and $U_{k_1 k_2 p}$ is the polariton-polariton scattering strength. The second term stands for the incoherent processes, which represents the summation of all the dissipative channels '$s$' associated with the dissipative operator (or jump operator) $\hat{J}_s$. By applying the technique discussed in [20], which describes a generic polariton system under an incoherent pumping and phonon-mediated scattering, for incoherent pumping, the original Hamiltonian in the Interaction picture reads as

$$\hat{H}_{pump} = \sum_k \left( \hat{a}_k(t) \hat{P}_k^\dagger(t) + \hat{a}_k^\dagger(t) \hat{P}_k(t) \right).$$ (4)

In Eq.(4), we define the pumping operator $\hat{P}_k(t) = \gamma_k \hat{d}_k e^{-iE_k t}$, where $\gamma_k$ is the decay rate of polaritons. The annihilation operator $\hat{d}_k$ corresponds to pumping reservoir particles, and their statistics follows the standard Bose distribution: $\langle \hat{d}_k^\dagger \hat{d}_k \rangle = \bar{n}_p(E_k)$. By tracing out the pumping part, we can get the specific jumping operators for pumping channel [20]:

$$\hat{J}_k^+ = \sqrt{\gamma_k \bar{n}_p(E_k)} \hat{a}_k^\dagger,$$ (5)

$$\hat{J}_k^- = \sqrt{\gamma_k [\bar{n}_p(E_k) + 1]} \hat{a}_k,$$ (6)

where the pumping intensity is described by $P = \gamma_k \bar{n}_p(E_k)$. Using a similar procedure, we find the jumping operators for the polariton-photon interaction channel,

$$\hat{J}_{k_1 k_2}^+ = \sqrt{\gamma_{k_1 k_2}^{ph} \bar{n}_{ph}(E_{k_1} - E_{k_2})} \hat{a}_{k_1}^\dagger \hat{a}_{k_2},$$ (7)

$$\hat{J}_{k_1 k_2}^- = \sqrt{\gamma_{k_1 k_2}^{ph} [\bar{n}_{ph}(E_{k_1} - E_{k_2}) + 1]} \hat{a}_{k_1} \hat{a}_{k_2}^\dagger.$$ (8)

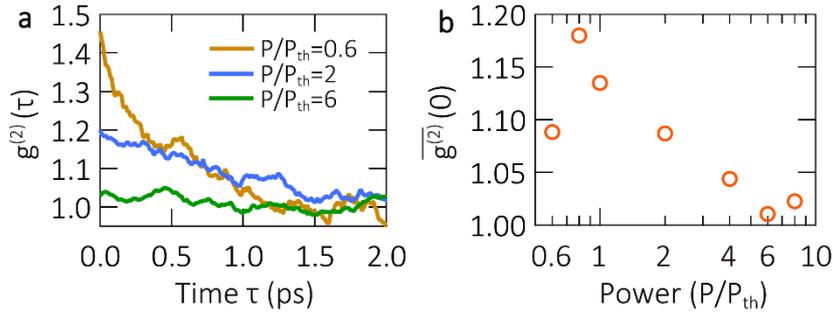

**Fig. 3 Simulation results for the second-order correlation function.** (a) Exemplary second-order correlation function $g^{(2)}(\tau)$ as a function of delay time $\tau$ at a temperature of 3.2 K. The pump power is P/P$_{th}$=0.6, 2, 6. (b) Averaged second-order correlation function at zero delay $\overline{g^{(2)}(0)}$ for different pump powers. It is calculated by averaging the original $g^{(2)}(\tau)$ data (as shown in panel a) over an integration period of 2 ps.

By employing the Monte Carlo wavefunction method integrated with the jumping operators defined above, we can calculate the time-averaged correlation function $\overline{g^{(2)}(\tau)}$ of the ground state. In our simulations, we apply the QuTiP package [28,29] to calculate $g^{(2)}(\tau)$, and to figure out $\overline{g^{(2)}(\tau)}$, we average the original $g^{(2)}(\tau)$ over a time period. The input parameters of simulations, including exciton energy and decay rate, cavity photon dispersion and decay rate, as well as coupling strength are all retrieved from the fit of coupled oscillator model in Fig. 1b. More details are presented in the Supplementary Material [22].

Typical simulation results of $g^{(2)}(\tau)$ at different pump powers are shown in Fig. 3(a). At zero delay time, $g^{(2)}(0)$ is calculated as 1.46 at power $P/P_{th}$=0.6. The function $g^{(2)}(\tau)$ quickly decays with delay time: it reduces to 1 at ~1.1 ps, and starts to fluctuate around unity afterward. For a power above the threshold ($P/P_{th}$=2), the value of $g^{(2)}(0)$ drops to 1.20. As the power is increased to $P/P_{th}$=6, we observe a further reduction of $g^{(2)}(0)$ down to 1.03. The $g^{(2)}(\tau)$ curve decays, and at some point, it turns out pinned to the unity level. This result depicts the power-dependent evolution of original $g^{(2)}(\tau)$, and the continuous reduction of $g^{(2)}(0)$ implies the growth of quantum-degenerate polariton lasing.

In Fig. 3(b), we show the time-averaged second-order correlation function at zero delay $\overline{g^{(2)}}(0)$ as a function of pump power. It is calculated here by integrating the $g^{(2)}(\tau)$ value over a time period of 2 ps, analogous to the experiment. At $P/P_{th}$=0.6, $\overline{g^{(2)}}(0)$ is ~1.09, which is much smaller than the $g^{(2)}(0)$ value (1.46) due to the averaging effect. With the increase of power, $\overline{g^{(2)}}(0)$ qualitatively follows the trend as observed in the experiment and approaches unity far above the threshold.

The calculated $\overline{g^{(2)}}(0)$ severely depends on the integration period. Thus, the corresponding values may not be exactly the same as in the measurements. In our case, we primarily focus on the pulse width of the excitation source as the dominating timescale. However, other factors such as the relaxation lifetime of polaritons (that is affected by phonons), may also impact the timescale of averaging.

To gain a deeper insight of the photon statistics of the polariton emission, we also study the second-order correlation function at different temperatures (corresponding to the experiment depicted in Fig. 2(d)). At 6.0 K, the value of $\overline{g^{(2)}}(0)$ (as a function of pump power) reproduces the trend at 3.2 K, exhibiting an increase from 1.01 to 1.02 below the threshold. Then, above the threshold, it is practically getting pinned to unity. As the temperature increases to 10.0 K, we find that the difference of $\overline{g^{(2)}}(0)$ value below and above the threshold power is getting reduced, which suggests that the phase transition is not as distinct as at lower temperatures. When the temperature is raised to 20.0 K, we cannot observe any $\overline{g^{(2)}}(0)$ variation along the input-output scan, which hints at the disappearance of a polariton phase transition due to the influence of phonons. The latter is further confirmed by the linearization of the intensity (as a function of pump-power curve) as the temperature is increased, as shown in the Supplemental material [22]. We interpret this temperature-dependent trend as a clear sign of temperature-induced depletion of the polaritonic ground state.

**CONCLUSION**

In summary, we studied the second-order coherence of quantum-degenerate polariton lasing in a cavity-integrated $MoSe_2$ monolayer. Experimentally, we observe distinct polariton dispersions and characteristic features of bosonic lasing including threshold-like behavior and spectral narrowing. The HBT measurements reveal the photon statistics of polariton emission, which transits from a photon bunching below the threshold to a coherent state above the threshold. We find the signatures of the polariton phase transition weakening with the increase of temperature, indicating thermally-induced ground-state depletion. Theoretically, by solving the Lindblad master equation, we model the second-order coherence of the dissipative polariton system on the quantum level and reproduce the time-averaged second-order correlation function of the ground state via the Monte Carlo wavefunction technique. Our findings elucidate the evolution of temporal coherence of a polariton lasing at different temperatures. These results pave the way towards quantum-degenerate lasers with improved working temperature, which can be used, e.g. in engineering polariton landscapes via introducing real-space potentials [12-14], and improving their feasibility for real-world applications.

The authors acknowledge support by Max Waldherr and Holger Suchomel in fixing the sample and growing the Bragg mirrors.
Financial support from the European Research Council within the project unLimit2D (Grant number 679288) and by the Niedersächsisches Ministerium für Wissenschaft und Kultur ("DyNano") is acknowledged. C.S. acknowledges support by the German Research Foundation (INST184/220-1 FUGG) S.H. acknowledges the Deutsche Forschungsgemeinschaft (DFG, German Research Foundation)–HO 5194/16-1 and INST 93/1007-1 LAGG. S.T. acknowledges funding from NSF DMR 1955889, DMR 1933214, and 1904716. S.T. also acknowledges DOE-SC0020653, DMR 2111812, and ECCS 2052527 for material development and integration. K.W. and T.T. acknowledge support from the JSPS KAKENHI (Grant Numbers 21H05233 and 23H02052) and World Premier International Research Center Initiative (WPI), MEXT, Japan. C. Anton-Solanas acknowledges the support from the Comunidad de Madrid fund "Atraccion de Talento, Mod. 1", Ref. 2020-T1/IND- 19785, grant no. PID2020113445-GB-I00 funded by the Ministerio de Ciencia e Innovación (10.13039/501100011033), and the grant ULTRA-BRIGHT from the Fundación Ramon-Areces in the "XXI Concurso Nacional para la adjudicación de Ayudas a la Investigación en Ciencias de la Vida y de la Materia". M.E. acknowledges funding by the University of Oldenburg through a Carl von Ossietzky Young Researchers' fellowship.

# Supplemental Material for the "Second-order temporal coherence of polariton lasers based on an atomically thin crystal in a microcavity"

Hangyong Shan[1,†], Jens-Christian Drawer[1,†], Meng Sun[2,†], Carlos Anton-Solanas[5], Martin Esmann[1], Kentaro Yumigeta[6], Kenji Watanabe[7], Takashi Taniguchi[8], Sefaattin Tongay[6], Sven Höfling[4], Ivan Savenko[3,9,10,*] and Christian Schneider[1,*]

[1]Institute of Physics, Carl von Ossietzky University, 26129 Oldenburg, Germany.
[2]Faculty of Science, Beijing University of Technology, 100124 Beijing, China
[3]Guangdong Technion Israel Institute of Technology (GTIIT), 241 Daxue Road, Shantou, Guangdong Province 515603, P.R. China
[4]Julius-Maximilians-Universität Würzburg, Physikalisches Institut and Würzburg-Dresden Cluster of Excellence ct.qmat, Lehrstuhl für Technische Physik, Am Hubland, 97074 Würzburg, Deutschland.
[5]Depto. de Física de Materiales, Instituto Nicolás Cabrera, Instituto de Física de la Materia Condensada, Universidad Autónoma de Madrid, 28049 Madrid, Spain.
[6]School for Engineering of Matter, Transport, and Energy, Arizona State University, Tempe, Arizona 85287, USA
[7]Research Center for Electronic and Optical Materials, National Institute for Materials Science, 1-1 Namiki, Tsukuba 305-0044, Japan
[8]Research Center for Materials Nanoarchitectonics, National Institute for Materials Science, 1-1 Namiki, Tsukuba 305-0044, Japan
[9]Technion -- Israel Institute of Technology, 32000 Haifa, Israel
[10]Guangdong Provincial Key Laboratory of Materials and Technologies for Energy Conversion, Guangdong Technion—Israel Institute of Technology, Guangdong 515063, China

[*]Corresponding author. Email: christian.schneider@uni-oldenburg.de, ivan.g.savenko@gmail.com
[†]These authors contributed equally: Hangyong Shan, Jens-Christian Drawer, Meng Sun

## S1. Supplied characterizations of the polariton sample and the setup

### Sample fabrication

The sample consists of an optical cavity containing an atomically thin layer of $MoSe_2$. The bottom DBR, grown by molecular beam epitaxy, has 24 mirror pairs of AlAs and $Al_{0.2}Ga_{0.8}As$ with thicknesses of 62 nm and 53 nm, respectively. The stop band is centered at 753 nm. On top of this DBR, there is a 52 nm thick spacer of AlAs containing a 4.75 nm thick GaAs quantum well (QW) 10 nm below the top surface. This spacer assembly is the lower half part of the optical cavity. It is followed by a layer of 3 nm thick GaAs onto which the $MoSe_2$ monolayer is transferred via the dry-gel stamping method. The monolayer flake is exfoliated from a bulk crystal grown by chemical vapor deposition. A layer of hexagonal boron nitride caps the flake. It is followed by a spin-coated ~90 nm thick PMMA buffer layer, which serves as the upper half part of the cavity. The sample is completed by placing a 8.5 mirror pair $SiO_2/TiO_2$ DBR on top via the micro-mechanical assembly technique. The stop band of the top DBR is centered at 750 nm.

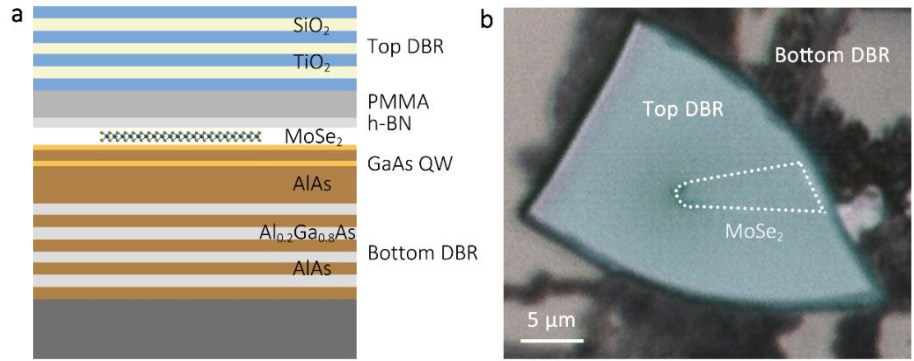

**Fig. S1 Sample structure and optical image. a** Illustration of the sample structure. The bottom DBR is grown by molecular beam epitaxy, and the top DBR is added by a micro-mechanical assembly technique. **b** Optical image of the sample. The top DBR has an anchor shape, with a size of ~15 μm × 25 μm. The monolayer $MoSe_2$ flake is outlined with dashed line.

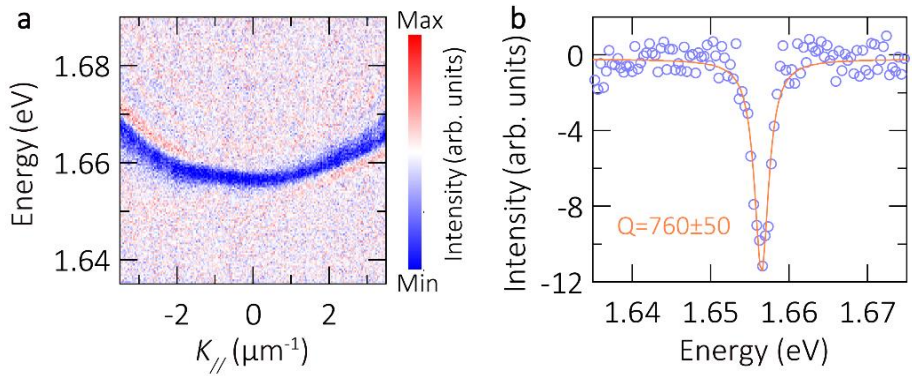

**Fig. S2 Dispersion relation of the uncoupled optical cavity. a** Angle-resolved white light reflectivity of the empty cavity away from the $MoSe_2$ monolayer at 3.2 K. **b** Reflectivity spectrum of uncoupled cavity at around $K_{//}=0$ μm⁻¹. It is fitted with a Lorentz function, the extracted quality factor is Q = 760±50.

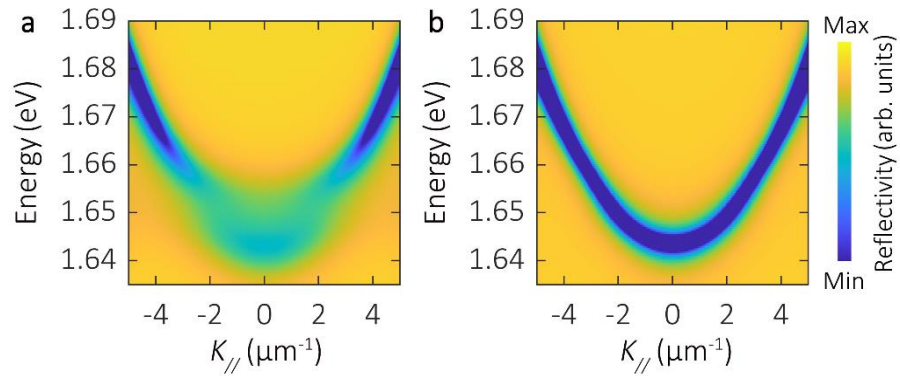

**Fig. S3 Simulation of dispersion relation with transfer matrix method. a** Simulated dispersion of polaritons. All structure parameters are extracted from experiments. The simulation result agrees with the measured dispersion relation in Fig. 1b. **b** Simulated dispersion of the case where the monolayer $MoSe_2$ is considered as a dielectric material. The imaginary part of the refractive index in $MoSe_2$ is set to zero, and the rest of the parameters are unaltered.

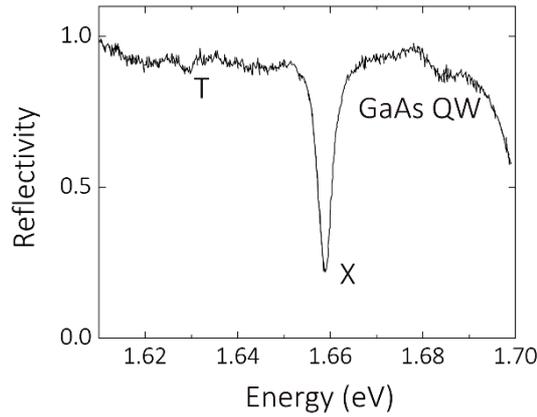

**Fig. S4** Reflectivity spectrum of the reference sample bottom DBRs/MoSe$_2$ monolayer/hBN. The GaAs QW shows a small dip. Instead, excitons (X) of MoSe$_2$ monolayer present remarkable absorption. Trions of MoSe$_2$ monolayer (T) are also observed with very small oscillator strength.

## Experimental setup

A standard back Fourier plane imaging setup is utilized to perform angle-resolved spectral measurements. In the excitation path, two different light sources are used, a Coherent Mira Optima 900-F mode-locked Ti:Sa laser operating at 740 nm with 2 ps pulse duration (76.2 MHz repetition rate), and a white light source (Thorlabs SLS301). All measurements are performed with the sample mounted on a motorized XYZ stage, model Attocube ANPx101/LT (for x, y) and ANPz51/RES/LT (for z), in a closed-cycle cryostat, model attoDRY1000, operating at 3.2 K. The first lens above the sample is a Thorlabs 354105-B with NA = 0.6. A back Fourier plane imaging technique with five lenses is used (Figure S5), offering the possibility of real space filtering in PL measurements. A 200 μm pinhole is mounted at the focal plane of real space in the setup (Figure S6), to conduct spatial filtering of polariton emission. Additionally, an extra lens can be flipped on to image the profile of laser and pinhole on a CMOS camera, and to align their relative position. Furthermore, to get rid of the strong laser background in the LP emission, the signal is spectrally filtered by a Semrock 790 nm VersaChrome Edge™ tunable long-pass filter and a Thorlabs FEL0750 long-pass filter. A charge-coupled device (CCD), model Andor iKon-M 934 Series, is attached to a spectrometer, model Andor Shamrock SR-500i. The CCD is used at highest sensitivity (50 kHz read-out rate, 4x pre-amplification, sensor temperature -85 °C), and the exposure time is set to 10 s.

For HBT measurements, the spectrally- and spatially-filtered emission signal (Figure S7) in free space is first recorded on a spectrometer to check if it is filtered rigorously. Next, it is directed into our HBT setup via a fiber launch system (Thorlabs KT120/M) with an aspheric lens (Thorlabs A240TM-B). A fiber (Thorlabs M42L05) is used to transmit the emission signal to a 50:50 fiber beam splitter (Thorlabs TM50R5F2A). Its two outputs are connected with two avalanche photodiodes (APDs) (Laser Components COUNT-T100-FC). We use a time tagger (Swabian Instruments, Time Tagger 20) to perform $g^{(2)}(\tau)$ correlation measurements. Temperature-dependent $g^{(2)}(\tau)$ measurements are conducted using a temperature controller (LakeShore Model 335).

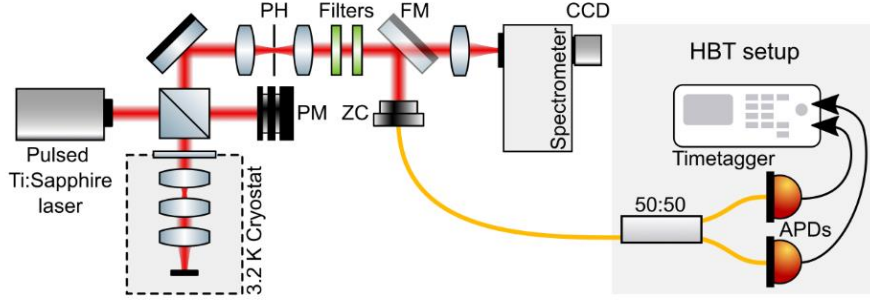

**Fig. S5 Schematic diagram of the setup.** Back Fourier plane imaging setup integrated with spatial filtering technique. The excitation source can be switched between mode-locked Ti:Sa laser and white light, to perform angle-resolved PL or reflectivity. The sample is loaded in a closed-cycle cryostat, operating at a temperature of 3.2 K. Because of the long length of the stick that is inserted into the cryostat, in our home-built optical microscope, three lenses are mounted on the stick to get clear optical images of the sample. In the collection path, a pinhole is mounted at the focal plane of real space. After the pinhole, we mount a flip lens: it is used to image the profile of the laser and pinhole on a CMOS camera and to align their relative position. When conducting angle-resolved measurements after alignment, this extra lens is flipped off the path. The signal is further filtered by two spectral filters, and the retained polariton emission is coupled into a fiber via a zoom collimator. The fiber-coupled signal is then input into a standard HBT setup to perform correlation measurements. PM: powermeter, PH: pinhole, ZC: zoom collimator, FM: flip mirror, CCD: charge-coupled device, APDs: avalanche photodiodes.

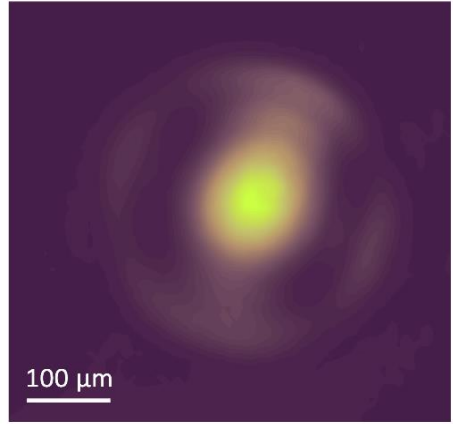

**Fig. S6 Spatial filtering of polariton emission using a 200 μm pinhole.** The pinhole is mounted at the focal plane of real space in the optical setup, and its profile is imaged on a CMOS camera, which is displayed as the outer circle in the image. The inner circle presents the excitation laser spot at the corresponding focal plane, which is aligned to the center of the pinhole. The spatial filtering technique effectively records the emission signal from the area of the pinhole and excludes the spectral influence of scattering at the edge of the top DBR.

Coupling strength in the fitting of coupled oscillator model

In the main text, we discuss the fit of coupled oscillator model, and here we calculate the coupling strength from mode splitting. We define detuning as $\delta = E_c - E_x$, and at zero detuning, the Rabi splitting $\Omega$ is given as: $\Omega = 2\sqrt{g^2 - \frac{1}{4}(\gamma_x - \gamma_c)^2}$ . This formula is valid for the splitting of polariton mode energies, but strictly speaking, it doesn't apply to the splitting peaks measured from experiments such as in reflectivity and PL spectra [1]. Specifically, the analytical formula to describe the splitting in PL is written as, $\Omega_{PL} = \sqrt{2\Omega\sqrt{\Omega^2 + 4\Gamma^2} - \Omega^2 - 4\Gamma^2}$ , where $\Omega$ is the Rabi splitting given above, and $\Gamma = \frac{1}{2}(\gamma_x + \gamma_c)$.

For reflectivity, as in our case, the formula is expressed as,

$\Omega_{Ref} = 2\sqrt{\sqrt{g^4(1 + \frac{2\gamma_x}{\gamma_c})^2 + 2g^2\gamma_x{}^2(1 + \frac{\gamma_x}{\gamma_c})} - 2g^2\frac{\gamma_x}{\gamma_c} - \gamma_x{}^2}$ , where $\Omega_{Ref}$ is $11\pm1$ meV, $2\gamma_c$ is 2.2 meV and $2\gamma_x$ is 2 meV. Solving this equation, we get a coupling strength g of $5.5\pm0.5$ meV.

## S2. Supplementary information on second-order temporal coherence measurements

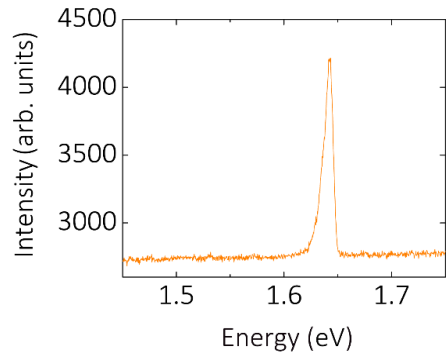

**Fig. S7 Emission signal inputting into the HBT setup.** Two tunable long-pass filters are used in combination with the spatial filtering technique, to obtain clean polariton emission spectra. The HBT setup measures correlations of the polariton ground state. To ensure the excitation laser is completely filtered out, the right side of the polariton emission is slightly cut, which is unavoidable in experiments under resonant excitation.

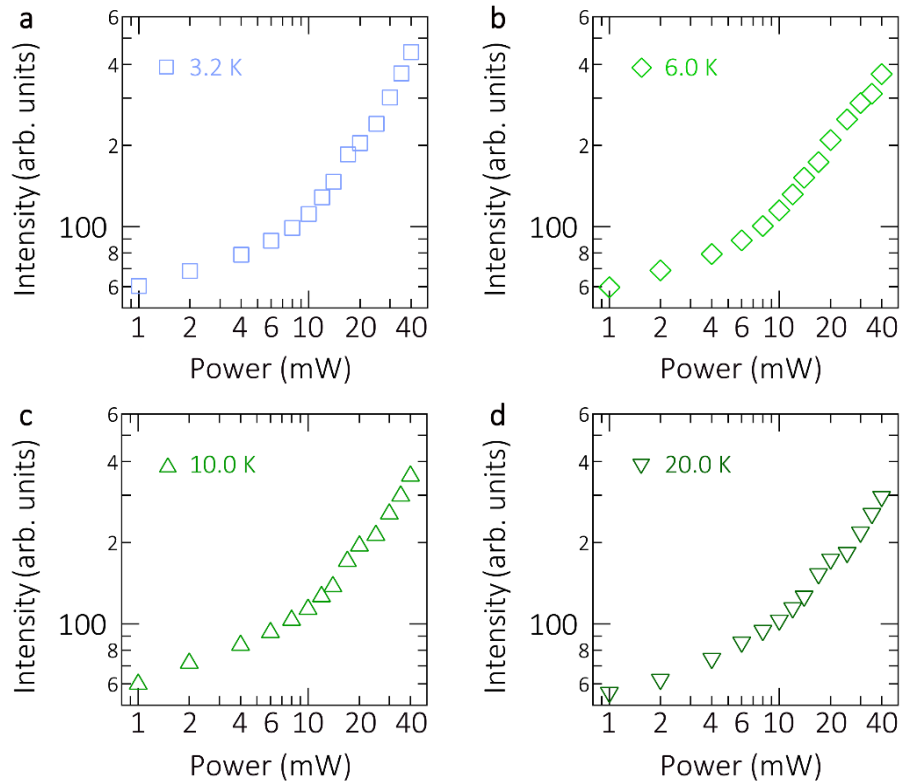

**Fig. S8 Intensity of uncorrelated peaks in $g^{(2)}(\tau)$ measurements at different temperatures. a – d** 3.2 K, 6.0 K, 10.0 K and 20.0 K. Data in panel a is reproduced from Fig. 2c. With the increase of temperature, the kink at threshold power is getting smoother, indicating the gradual disappearance of lasing behavior.

## S3. Supplementary information on theoretical simulations

In our simulations, we consider the microcavity has cylinder symmetry; For eigenstates of lower polariton $E_{LP}(k_{\parallel})$ expressed in the main text, we extracted the dispersion of cavity photons from the fit of coupled oscillator model in Fig. 1(b), given as $E(k_{\parallel}) = 1.6465 + 0.0016\, k_{\parallel}^2$ (eV); the effective mass of exciton is treated as infinity as usual; the decay rate of the cavity photon and exciton are $2\hbar\gamma_c = 2.2\,meV$ and $2\hbar\gamma_x = 2.0\,meV$, respectively; due to the symmetry, we choose five states equally distributed between $k_{\parallel} \in [0,4]\,\mu m^{-1}$ in our simulation and the maximum number of excitations is $N_{max} = 12$ for the ground state ($k_{\parallel} = 0\,\mu m^{-1}$) and $N_{max} = 4$ for the other states; for polariton-polariton interaction, we disregard any coupling between different modes and assume the interaction strength is a constant for the different modes as $U/N_{max} = 1.5\,meV$; at last, we define the effective polariton-phonon scattering strength as a constant for all the different modes $\hbar\gamma_{k_1 k_2}^{ph} \equiv \hbar\gamma^{ph} = 1.0\,meV$.

With this setting, we calculate the averaged second-order correlation function of the ground state. In the calculations, we use the quantum-jump approach [2], which simulates the Lindblad equations for the single-particle density-matrix [3,4]. To begin with, we first let the system evolves for $2ps$ ($4ps$) with 500 realizations above (below) the threshold. Then, for each realization $g^{(2)}(\tau)$ is averaged by another 500 trajectories for $2ps$. Thus, the result of $g^{(2)}(\tau)$ is averaged over $500 \times 500$ trajectories in total.